\documentclass[preprint,12pt]{elsarticle}

\usepackage{amssymb}
\usepackage{soul}
\usepackage{amsmath}
\usepackage{subcaption}
\usepackage{xcolor}
\usepackage{placeins} 

\journal{Acta Astronautica}

\begin{document}

\begin{frontmatter}




\title{Performance Bounds for Cooperative Localisation \\ in the Starlink Network}
\author[inst1,inst2]{Calum Turner\corref{cor1}}

\affiliation[inst1]{organization={Faculty of EEMCS},
            addressline={Delft University of Technology}, 
            city={Delft},
            postcode={2628}, 
            country={The Netherlands}}

\author[inst1]{Raj Thilak Rajan}

\affiliation[inst2]{organization={Space Advanced Concepts Laboratory},
            addressline={ISAE-SUPAERO}, 
            city={Toulouse},
            postcode={31400}, 
            country={France}}

\tnotetext[t1]{Dr. R.T.Rajan is partially funded partially by the Dutch-PIPP (Partnerships for Space
Instruments Applications Preparatory Programme), funded by NWO (Netherlands Organisation for Scientific Research) and
NSO (Netherlands Space Office).}

\begin{abstract}
Mega-constellations in Low Earth Orbit have the potential to revolutionise worldwide internet access. The concomitant potential of these mega-constellations to impact space sustainability, however, has prompted concern from space actors as well as provoking concern in the ground-based astronomy community. Increasing the knowledge of the orbital state of satellites in mega-constellations improves space situations awareness, reducing the need for collision avoidance manoeuvres and allowing astronomers to prepare better observational mitigation strategies. In this paper, we create a model of Phase 1 of Starlink, one of the more well-studied mega-constellations, and investigate the potential of cooperative localisation using time-of-arrival measurements from the optical inter-satellite links in the constellation. To this end, we study the performance of any unbiased estimator for localisation, by calculating the instantaneous Cram$\acute{\text{e}}$r-Rao bound for two situations; one in which inter-satellite measurements and measurements from ground stations were considered, and one in which only relative navigation from inter-satellite measurements were considered. Our results show that localisation determined from a combination of inter-satellite measurements and ground stations can have at best an an average RMSE of approximately $10.15$ metres over the majority of a satellite's orbit. Relative localisation using only inter-satellite measurements has a slightly poorer performance with an average RMSE of $10.68$ metres. The results show that both anchored and anchorless inter-satellite cooperative localisation are dependent on the constellation's geometry and the characteristics of the inter-satellite links, both of which could inform the use of relative navigation in large satellite constellations in future.  
\end{abstract}



\begin{keyword}
Starlink \sep Localisation \sep Cram$\acute{\text{e}}$r-Rao Bound \sep Swarms \sep Machine learning \sep Signal Processing

\end{keyword}

\end{frontmatter}



\section{Introduction}
Improving the knowledge of the orbital state of satellites is a crucial element of space sustainability, particularly space traffic management and space situational awareness. Within distributed science missions, accurate relative positions are vital for science missions incorporating distributed \cite{APIS} or  interferometric measurements \cite{OLFAR}. In a connected network of satellites, the inter-satellite links allow cooperative localisation to be performed based on satellite-to-satellite measurements. This provides additional information to operators seeking to improve space situational awareness, reduces dependency on ground stations, and provides a redundant method of localising satellites to any guidance, navigation, and control hardware on board. The improved knowledge of orbital position can also benefit space sustainability beyond space traffic management. Knowing the precise location of satellites allows astronomers to time their observations to avoid satellite trails, which would otherwise saturate the sensitive detectors in large telescopes. The potential use of inter-satellite measurements for autonomous navigation has received growing academic attention in the last several years, with research investigating the performance of autonomous navigation using laser inter-satellite links in a variety of Earth orbits \cite{dave2020autonomous} and investigating the use of laser inter-satellite links for precise orbit determination in constellations of up to 192 satellites in LEO \cite{li2019leo}. Other studies have focused on the use of laser inter-satellite links for both orbit and clock corrections determination \cite{kur2021application}.

In this paper, we model the performance of the position estimation of satellites within a megaconstellation in Low Earth Orbit (LEO) e.g., the Starlink network \cite{Starlink_website}. First, we create a spatial model of Phase-1 of the Starlink network, which represents the locations of the satellites in the network. The Cram$\acute{\text{e}}$r-Rao Bound is then used to establish a lower bound on the achievable localisation performance considering scenarios with (a) only inter-satellite measurements and (b) the inter-satellite measurements augmented with measurements from ground stations. We choose to model Starlink among the possible mega-constellations since it is a relatively well-studied constellation that will employ optical inter-satellite links. Starlink is a large LEO constellation, in which thousands of satellites exchange data to provide low-latency internet \cite{bhattacherjee_network_2019}, making Starlink a network of intercommunicating satellites collectively operating as a distributed system. Starlink is also an interesting case study as it has been noted as contributing to concerns about space sustainability \cite{boley_satellite_2021} and interference with ground-based astronomy \cite{hainaut_impact_2020} \cite{mcdowell_low_2020}. Our choice of Starlink as case was also influenced by previous research addressing the inter-satellite links between Starlink satellites and the resulting network topology \cite{bhattacherjee_network_2019} \cite{chaudhry_laser_2021}.

The Starlink mega-constellation is currently under construction by SpaceX (Space Exploration Technologies) in LEO. The constellation is will eventually require several thousand satellites for global coverage. Launches of operational satellites began in 2019, and as of 1$^{st}$January 2022 roughly 40\% of all active satellites in orbit belonged to the Starlink constellation \cite{union_concerned_sceintists_satellite}. January 2022 was the most recent update of the Union of Concerned Scientists' Satellite Database, but at the time of writing more than 2000 Starlink satellites are in orbit.

\textit{Outline:} The outline of this paper is as follows. We introduce the Starlink network model in Section \ref{Methods}, as well as discussing the topology of the Starlink intersatellite network and the location and visibility of Starlink ground stations. Section \ref{calculation} describes Starlink as a cooperative localisation problem for a wireless network and introduces the Cram$\acute{\text{e}}$r-Rao Bound and the formulae used to calculate our results. Our results are presented in Section \ref{section:Results} and the paper concludes with a discussion and summary of our results as well as prospects for further study. 

\section{Starlink network}
\label{Methods}

In this Section we discuss the simulation to obtain the locations of the satellites in the Starlink network. We also discuss the inter-satellite network topology we assumed for Starlink i.e., which satellites communicate with one another in the constellation. Furthermore, we also describe the location of the Starlink ground stations and how the visibility of the satellites are calculated from each ground station.

\subsection{Modelling Starlink} \label{section:starlink_model}
To know the true locations of the Starlink satellites, a model of the constellation was created using \textsc{Python}. The swarm consisted of 1584 satellites in Low Earth Orbit at an altitude of 550 km, corresponding to Phase 1 of the Starlink constellation. The details of this constellation design were based on the information in an FCC filing dated April 17, 2020 \cite{wiltshire_application_2020}, and the parameters of this orbit are presented in Table~\ref{tab:starlink_parameters}. The satellites' orbits —which were assumed to be circular— were propagated using Poliastro, an open-source \textsc{Python} library for astrodynamics \cite{juan_luis_cano_rodriguez_2021_5035326}. The J$_2$ effect was calculated for each satellite but aerodynamic drag was found to have a negligible effect on the satellite positions over the course of one orbit and was therefore omitted. Following the methodology in \cite{chaudhry_laser_2021}, each Starlink satellite was given a unique identifier with the format \textsc{sXXYYY} where \textsc{XX} is plane number and \textsc{YYY} is the satellite number in base 10. For example, the first satellite in the first plane has the identifier \textsc{s01001} and has initial position $[a, 0, 0]$ where $a$ is the semi-major axis of the orbit. The time-varying positions, velocities, and orbital elements of all 1584 satellites in the simulated Starlink constellation can be accessed at \cite{dataport}.


\begin{table}[!ht]
\renewcommand{\arraystretch}{1.3}
\caption{\bf Orbital Parameters of the Starlink network}
\label{tab:starlink_parameters}
\centering
 \begin{tabular}{|l||l|}
 \hline
  \textbf{Parameter} & \textbf{Value} \\
  \hline
    Altitude & 550 km \\ 
    Number of Planes & 72  \\
    Satellites per Plane & 22  \\
    Inclination \textit{i} & 53$^\circ$\\
    Orbital Period \textit{T} & 1.59 hours \\
    Total Satellites & 1584 \\
   \hline
\end{tabular}
\end{table}

\subsection{Network Topology}
\label{section:starlink_model_topology}
Starlink satellites will eventually be connected with optical inter-satellite links, allowing the system to transmit information and carry internet traffic, however the openly available information about these inter-satellite links and the corresponding subsystems is sparse. To determine which links were possible within the inter-satellite network, we considered three network constraints: visibility, range, and hardware limitations. Each constraint is described in detail below.

\begin{itemize}
    \item{\textbf{Range:} The distance between satellites determines whether or not they can establish a link.}
    \item{\textbf{Visibility:} The visibility of a satellite, which is also referred to as Line-Of-Sight (LOS), indicates if the satellite can receive the transmitted signal from another satellite without reflection or occlusion of the signal. In LEO, the presence of a central body with the radius of the Earth ($R_E=6371$ km) and the height of the ionosphere ($h=80$ km) place an upper limit on the range of a LOS links in LEO. Simple geometry gives a maximum link length of $x = 2 \sqrt{(R_E + a)^2 - (R_E+h)^2} = 5016$ km at an altitude of $a=550$ km \cite{bhattacherjee_network_2019,chaudhry_laser_2021,dave2020autonomous}.}
    \item{\textbf{Hardware:} The range and the LOS place physical constraints on potential links, but the design of the satellites themselves also affect how many links are feasible. In our example, the number of laser links that each satellite can support is limited in practice by the number of optical heads on each satellite.}
\end{itemize}

\begin{figure*}[ht]
    \centering
    \begin{subfigure}[t]{0.5\textwidth}
        \centering
        \includegraphics[width=\textwidth]{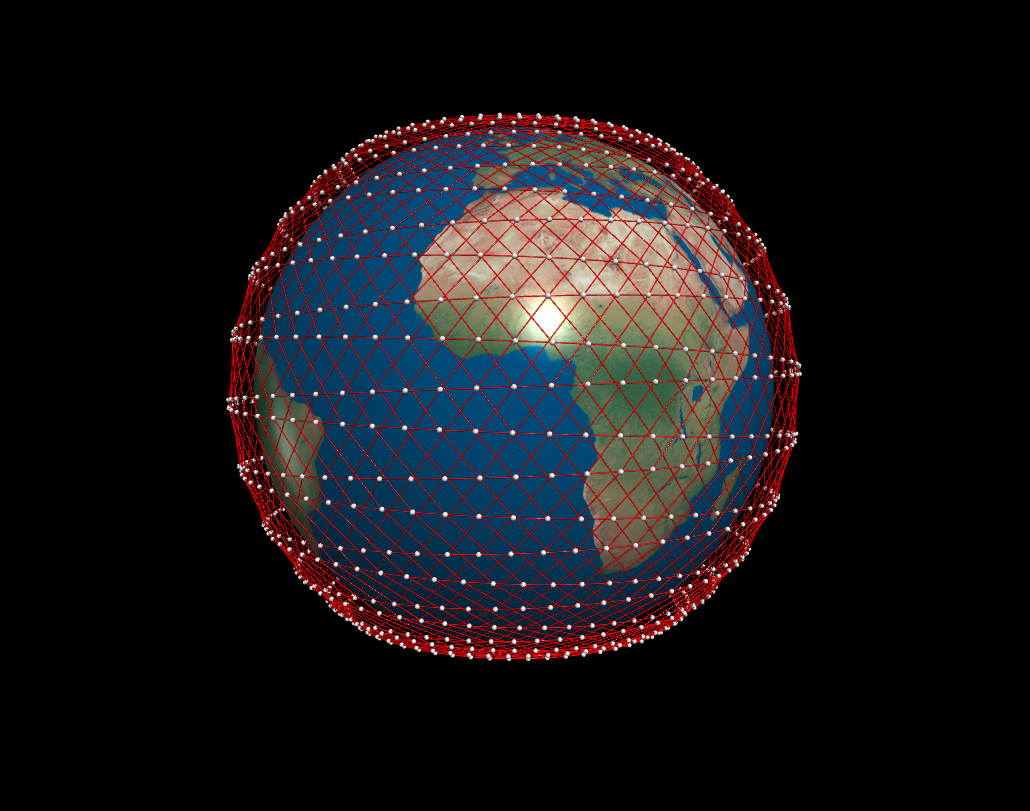}
        \caption{Full Starlink Network}
        \label{fig:network_A}
    \end{subfigure}%
    ~ 
    \begin{subfigure}[t]{0.5\textwidth}
        \centering
        \includegraphics[width=0.9\textwidth]{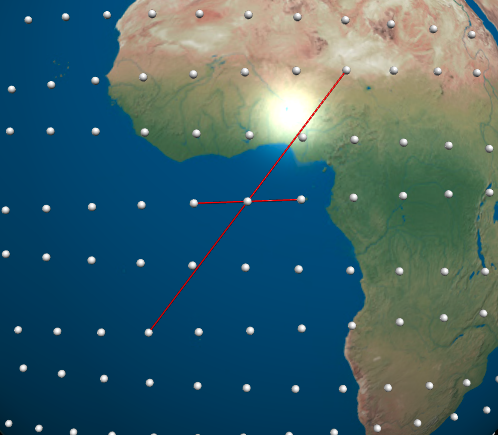}
        \caption{Connections for one satellite}
        \label{fig:network_B}
    \end{subfigure}
    \caption{Full network of the Starlink mega-constellation assuming a +grid topology. Figure \ref{fig:network_A} shows the full network, which is made from repeated patterns of the form shown in \ref{fig:network_B}. This network assumes that each satellite in limited to four connections due to hardware constraints. The figure also shows the positions of the satellites in the constellation in white, which are distributed evenly across 72 orbital planes with 22 satellites in each plane.}
    \label{fig:network}
\end{figure*}

After analysing the time-varying links possible in the Starlink constellation sing a similar analysis to that presented in \cite{chaudhry_laser_2021},  we have established that the Starlink satellites are capable of connecting to up to roughly 40 other satellites under the physical constraints of of visibility. In practice, however, the hardware on the satellite limits the number of possible inter-satellite links. We assumed that each Starlink satellite could connect to 4 nearby satellite, and following \cite{bhattacherjee_network_2019} we assumed a "+grid" network topology in which satellites are connected to two in the same orbital plane and two in neighbouring planes. The resulting network is shown in Figure \ref{fig:network_A} and the pattern of inter-satellite links for a single satellite is shown in Figure \ref{fig:network_B}.

\subsection{Ground Stations}
\begin{figure}[t]
    \centering
    \includegraphics[width=0.95\textwidth]{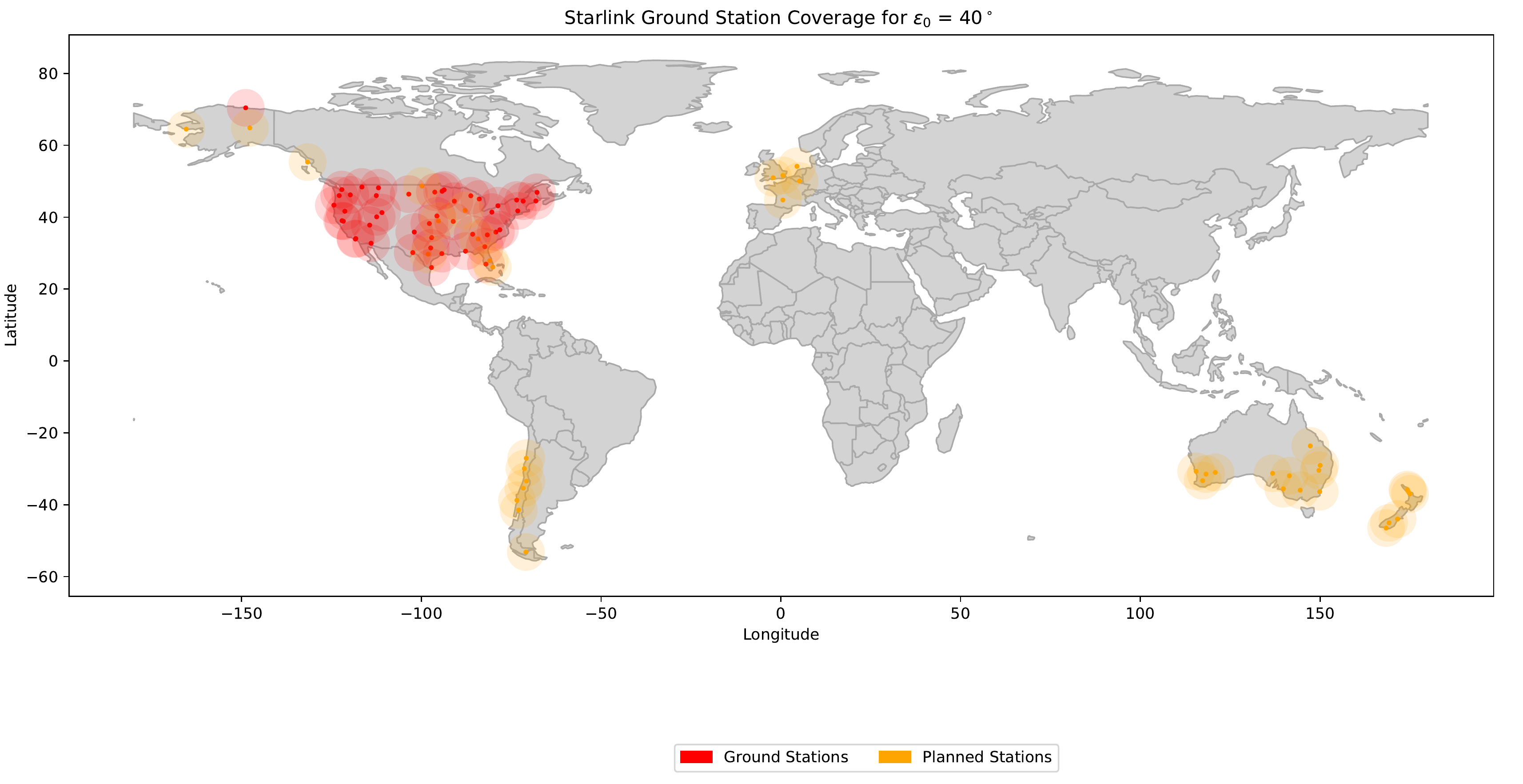}
    \caption[Starlink Anchor Locations]{The locations of 87 planned and active Starlink ground station and their coverage assuming $\epsilon_0$ = 40$^\circ$ \cite{cakaj_parameters_2021}. The locations of the ground stations are substantiated by filings with the Federal Communications Commission \cite{FCC} in the USA, República De Chile Ministerio De Transportes Y Telecomunicaciones Subsecretaría De Telecomunicaciones \cite{Chile_subsecretaritelecomunicaciones} in Chile, Autorité de Régulation des Communications Électroniques et des Postes \cite{arcep_nodate} in France, the Office of Communications in the UK \cite{arcep_nodate}, the Australian Communications and Media Authority \cite{Australia_authority_home_nodate} and the Ministry of Business, Innovation \& Employment, Radio Spectrum Management \cite{zealand_welcome_nodate} in New Zealand. All latitudes and longitudes are approximate (correct at the town/city level). Setting $\epsilon_0$ = 40$^\circ$ gives a visibility range of $d$ = 812 km.}
    \label{fig:starlink_anchor_positions}
\end{figure}

\begin{figure}[t]
    \centering
    \includegraphics[width=0.95\textwidth]{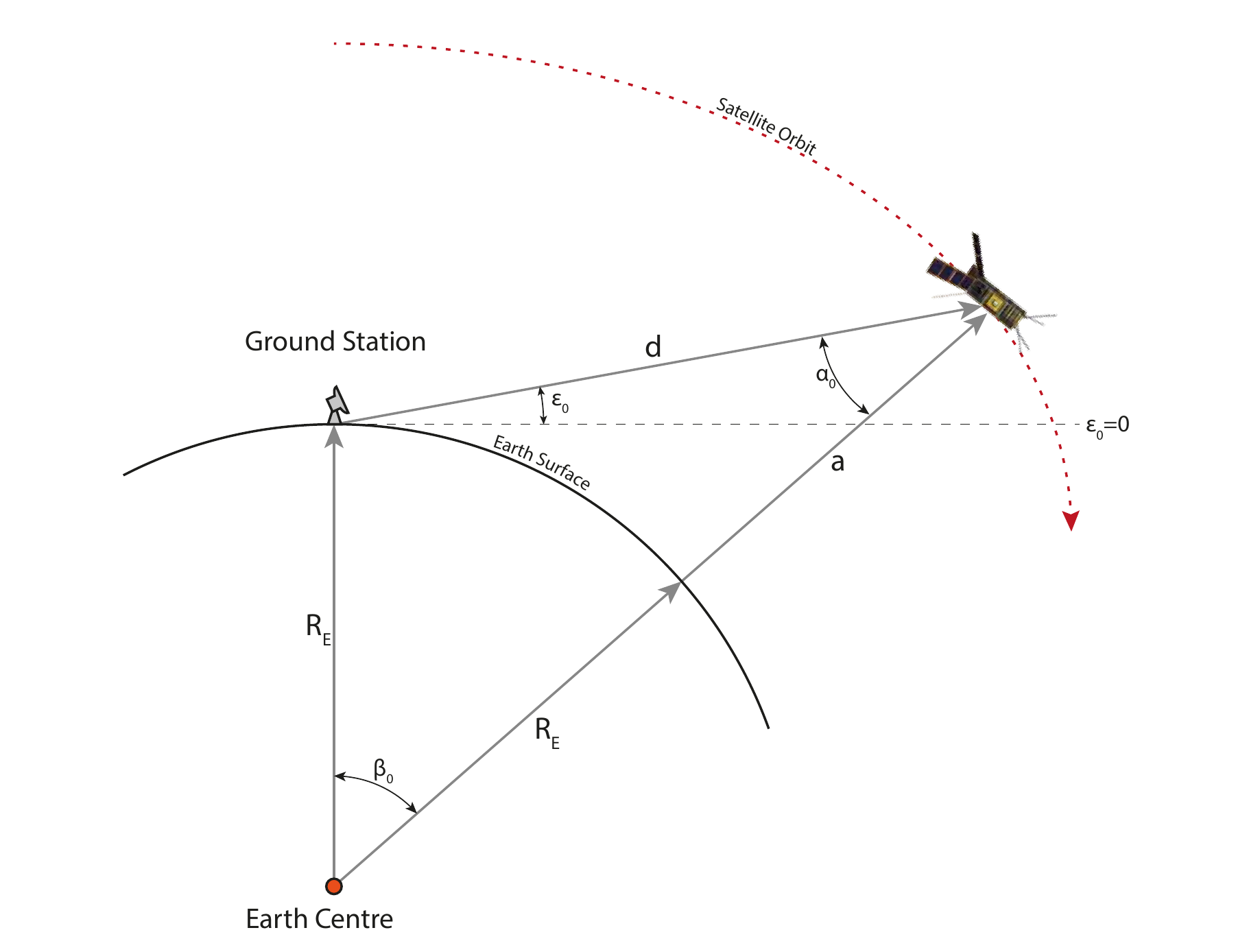}
    \caption[Geometric Set-up for Ground Station Visibility]{A diagram of the geometric set-up used to calculate the visibility of satellites from a ground station. The satellite has an altitude of $a$ and makes an angle of $\epsilon_0$ with the ground station's local horizon. The maximum distance at which the satellite is visible can be found by determining $d$. Diagram adapted from \cite{cakaj_range_2011}.}
    \label{fig:anchor_visibility}
\end{figure}

To calculate the CRB on the positions of the satellite locations in the Starlink network also requires the location of the megaconstellation's ground stations to be known. This allows the satellites visible from any given ground stations at any given time to be calculated. Figure \ref{fig:starlink_anchor_positions} shows the location of 87 planned or active Starlink ground stations as based on regulatory filings in the USA, Chile, UK, France, Australia, and New Zealand. We calculated the visibility of satellites from the ground stations shown in Figure \ref{fig:starlink_anchor_positions} using the equations detailed in \cite{cakaj_range_2011}. The geometric set-up for determining the maximum distance between a satellite and a ground station is shown in Figure \ref{fig:anchor_visibility}, and starting from this geometric set-up the maximum distance between a ground station and a satellite can easily be derived from the cosine law for triangles, which is given by \begin{equation}
    r^2 = R_E^2 + d^s - 2 \cdot R_E \cdot d \cdot \cos(\frac{\pi}{2}+\epsilon_0)
\end{equation} where $r$ is the orbital radius ($r=R_E+a$), $R_E$ is the radius of the Earth, $d$ denotes the distance between a satellite and the ground station, $a$ is the altitude of the satellite, and  $\epsilon_0$ is the elevation of the satellite above the ground station's local horizon. Rearranging this equation by using the quadratic equation and further simplification yields \begin{equation}
   \frac{2 \cdot R_E \cdot \cos(90+\epsilon_0) \pm \sqrt{4 \cdot R_E^2 \cdot \cos^2(\frac{\pi}{2} + \epsilon_0) - 4 \cdot (R_E^2 + r^2)}}{2}
\end{equation}

Applying simple trigonometric identities ($\cos^2\theta - \sin^2\theta = 1$, $\cos\theta = \sin(\frac{\pi}{2}-\theta)$, and $\sin(-\theta)=\sin\theta$, we have \begin{equation}
    d = R_E \left[ \sqrt{\left( \frac{r}{R_E} \right)^2-\cos^2 \epsilon_0} -\sin \epsilon_0\right]
\end{equation}

Finally substituting $r=R_E+a$ gives the maximum distance to a satellite at an elevation angle of $\epsilon_0$ above the ground station's horizon:

\begin{equation}
\label{eq:ground_station_range}
    d = R_E \left[ \sqrt{\left( \frac{a+R_E}{R_E} \right)^2-\cos^2 \epsilon_0} -\sin \epsilon_0\right]
\end{equation}

To find the maximum possible distance at which a satellite would be able to be tracked by a ground station, we set $\epsilon_0 = 0$, giving $\cos^2 \epsilon_0 = 1$ and $\sin \epsilon_0 = 0$. The equation simplifies to:

\begin{equation}
    d = R_E \left[ \sqrt{\left( \frac{a+R_E}{R_E} \right)^2-1} \right]
\end{equation}

Plugging in the values of $a$=550km and $R_E$ = 6371 km gives a maximum distance of 2704 km, in line with the results presented in \cite{cakaj_range_2011}. In practice, barriers such as hills, forests, or buildings mean that satellite operators often define a safe margin for $\epsilon_0$ that avoids these barriers \cite{cakaj_range_2011}. This value ranges from $\epsilon_0 = $ 0$^\circ$ to 30$^\circ$ \cite{cakaj_rigorous_2007,cakaj_simulation_2010}.

Using these ground station locations in to the Starlink model shows that between 114 and 135 satellites are connected at any given time over the course of an orbit, \textit{i.e.} between 7.8\% and 9.1\% of the total number of satellites\footnote{Note that these percentages will increase as more ground stations are added to the Starlink network}. The relatively low proportion of connected satellites despite 87 ground stations can be attributed to the relatively conservative value of $\epsilon_0$ = 40$^\circ$ we adopted based on \cite{cakaj_parameters_2021}. There are an average $341^{+18}_{-22}$ satellite-to-ground-station connections at any given time, \textit{i.e.} many satellites are connected to multiple ground stations. This arises as the coverage of some ground stations overlap, as shown in Figure \ref{fig:starlink_anchor_positions}.

With the Starlink model developed, the connections between the satellites determined, and the locations of the ground stations established, it is possible to determine the Cram$\acute{\text{e}}$r-Rao Bound for cooperative navigation in Starlink. First, however, we will introduce the cooperative localisation problem.

\section{Cooperative localisation}
\label{calculation}

In this section, we aim to understand the performance of potential localisation algorithms for estimating the positions of the Starlink network. To achieve this goal, we calculate the lower bound on the variance of potential unbiased estimators for localisation, using the Cram$\acute{\text{e}}$r-Rao Bound (CRB) \cite{patwari_locating_2005}.


\subsection{Problem Statement} The 3-dimensional cooperative sensor location estimation problem can be stated as follows. Consider $n$ nodes with unknown locations and $m$ reference nodes (also referred to as anchor nodes) with exactly known locations. The problem is to estimate the $3n$ unknown coordinates $\mathbf{\theta} = [\mathbf{\theta}_x,\mathbf{\theta}_y,\mathbf{\theta}_z]$, where \begin{equation*}
    \mathbf{\theta}_x = [x_1,x_2,...,x_n] \text{  ,  }\mathbf{\theta}_y = [y_1,y_2,...,y_n] \text{  ,  }\mathbf{\theta}_z = [z_1,z_2,...,z_n] 
\end{equation*} given the location of the reference nodes, $[x_{n+1},...,x_{m},y_{n+1},...,y_{m},z_{n+1},...,z_{m}]$ and a collection of distance measurements between the nodes. When cooperatively localising nodes, the measurements between nodes can capture various properties, including  the propagation time of the signals, the strength of received signals, or the angles from which signals are received \cite{patwari_locating_2005}. In this paper we considered Time-of-Arrival (ToA) measurements, in which inter-node distances are calculated by dividing the time of propagation of a signal by the velocity of propagation. For radio or optical signals propagating in a vacuum, this is simply $\frac{\Delta t}{c}$ where $c$ is the speed of light and $\Delta t$ is the time of flight. This method requires either the internal clocks of nodes and their biases to be known or estimated, or the clocks to all be synchronised. However, in \cite{rajan_joint_2015} it was shown that it is always possible to synchronise the clocks of a mobile anchorless network, subject to the constraint that each node has at least one 2-way connection to another node in the network. For our model of connected Starlink satellites, this implies that the satellite clocks can always be considered to be synchronised.

Treating the $n=1584$ Starlink satellites as the unknown nodes and the $m=87$ ground stations as nodes with known location, and considering the network topology described in Section \ref{section:starlink_model_topology}, it is possible to frame cooperative localisation in Starlink as a cooperative localisation problem for a wireless sensor network and to apply the Cram$\acute{\text{e}}$r-Rao Bound.

\subsection{Cram$\acute{\text{e}}$r-Rao Bound}
\label{section:crb}
The Cram$\acute{\text{e}}$r-Rao Bound (CRB) provides a lower bound on the variance that can be achieved by any unbiased estimator \cite{kay_fundamentals_1993} \cite{trees_detection_2004}. Essentially, the CRB is one of many performance bounds can be used to determine the 'best case' performance of an estimator at a given point with given information and using a given technique. The bound is affected by a number of parameters, including: \begin{itemize}
    \item{The number of sensors with unknown locations (nodes) and the number of sensors with known locations (anchors)}
    \item{Sensor geometry}
    \item{Dimensionality (3D or 2D localisation)}
    \item{Type of measurement (\textit{i.e.} Received Signal Strength (RSS), Time of Arrival (ToA), or Angle of Arrival (AoA)}
    \item{Link parameters}
    \item{Network topology (which pairs of sensors make measurements)}
\end{itemize}

In practice, the Cram$\acute{\text{e}}$r-Rao Bound can be determined by inverting the Fisher Information Matrix, \textbf{F}. Inverting the Fisher matrix $\textbf{F}$ gives the CRB matrix whose diagonals are the best achievable $x$,$y$, and $z$ location variances. To generate a single figure of merit, the calculated the square Root of the Cram$\acute{\text{e}}$r-Rao Bound (RCRB) for the $x$,$y$, and $z$ location using \begin{equation}
    \label{eq:CRB_RMS}
    \text{RCRB} 
    = \left(\left(1/n\right)\text{tr}(\textbf{F}^{-1})\right)^{1/2}
\end{equation} where $\text{tr}(\textbf{F}^{-1})$ is the trace of the inverse Fisher matrix and $n=3$ for the 3-by-3 Fisher Information Matrix.

\subsection{Anchored Cram$\acute{\text{e}}$r-Rao Bound}
\label{derivation:anchored}

With the satellite positions, network topology, and ground station locations defined for our Starlink model, it was possible to calculate the Fisher Information Matrix for each Starlink satellite. We calculated an individual Fisher matrix for each satellite at each timestep to reduce the run-time of the simulation by reducing the size of the computationally intensive matrix inversion, and also because calculating individual Fisher matrices is more appropriate for a distributed satellite system. The 3-by-3 Fisher Information Matrix \textbf{F} for satellite \textit{i} with position $\mathbf{x} =[x_i,y_i,z_i]$ is given by:

\begin{equation}
    \label{eq:Fisher_anchored}
    \textbf{F} = \gamma
    \begin{bmatrix}
    \textbf{F}_{xx} & \textbf{F}_{xy} & \textbf{F}_{xz}\\
    \textbf{F}_{xy}^T & \textbf{F}_{yy} & \textbf{F}_{yz} \\
    \textbf{F}_{xz}^T & \textbf{F}_{yz}^T & \textbf{F}_{zz} 
    \end{bmatrix}
\end{equation} where \begin{eqnarray}
\textbf{F}_{xx}&=& \sum_{j \in H(i)}(x_i-x_j)^2/d^s_{ij} \\ 
\textbf{F}_{yy}&=& \sum_{j \in H(i)}(y_i-y_j)^2/d^s_{ij} \\
\textbf{F}_{zz}&=& \sum_{j \in H(i)}(z_i-z_j)^2/d^s_{ij} \\
\textbf{F}_{xy}&=& \sum_{j \in H(i)}(x_i-x_j)(y_i-y_j)/d^s_{ij} \\
\textbf{F}_{xz}&=& \sum_{j \in H(i)}(x_i-x_j)(z_i-z_j)/d^s_{ij} \\
\textbf{F}_{yz}&=& \sum_{j \in H(i)}(y_i-y_j)(z_i-z_j)/d^s_{ij}
\end{eqnarray} where $H(i)$ is the set of nodes with which satellite $i$ can communicate and consists of the four connected satellites in the +grid network as well as any ground stations within range. $d_j$ is the distance between the Starlink satellite $i$ and connected node $j$ with position $[x_j,y_j,z_j]$. $s$ is an exponent dependent on measurement type, with $s=2$ for ToA, and $\gamma$ is a channel constant determined by the type of measurement which for ToA measurements is given by \begin{equation}
    \label{eq:gamma}
     \gamma = \frac{1}{(v_p \sigma_T)^2}
\end{equation} where $v_p$ is the propagation velocity of the signal and $\sigma_T$ is the standard deviation of the ToA measurements. The  CRB for anchored localisation of the satellite network is given by substituting (\ref{eq:Fisher_anchored}) in (\ref{eq:CRB_RMS}).

\subsection{Anchorless Cram$\acute{\text{e}}$r-Rao Bound}
\label{derivation:anchorless}
We also calculated the anchorless CRB, the localisation performance of the Starlink model calculated without including measurements from ground stations. It is important to note that this gives only the relative localisation performance of the Starlink satellites and not their absolute localisation performance. Using  the expressions and notation from  \cite{rajan_26}, the Fisher Information Matrix for the relative positions of all satellites, $\textbf{F}_{rel}$, is given by: \begin{equation}
    \textbf{F}_{rel} = \gamma \left[ \frac{\partial \mathbf{a}_x (\mathbf{\Phi}_x) }{\partial \mathbf{\Phi}_x^T}\right]^T \left[ \frac{\partial \mathbf{a}_x (\mathbf{\Phi}_x) }{\partial \mathbf{\Phi}_x^T}\right]
\end{equation} where $\mathbf{a}_x (\mathbf{\Phi}_x)$ is the set of measured distances between the n satellites and $\mathbf{\Phi}$ is the set of positions $\mathbf{\Phi} \triangleq \left[ \mathbf{x}_1^T,\mathbf{x}_2^T,\cdots,\mathbf{x}_n^T\right]^T \in \mathbb{R}^{3n \times 1}$ for the $n$ satellites in $3$D, where each $\mathbf{x}_i = [x_i,y_i,z_i]$, and $\gamma$ is defined by  (\ref{eq:gamma}). The Jacobian for the full constellation of $n$ satellites takes the form: \begin{equation}
    \frac{\partial \mathbf{a}_x (\mathbf{\Phi}_x) }{\partial \mathbf{\Phi}_x^T} = \left[ \frac{\partial \mathbf{a}_x (\mathbf{\Phi}_x) }{\partial \mathbf{x}_1^T},\frac{\partial \mathbf{a}_x (\mathbf{\Phi}_x) }{\partial \mathbf{x}_2^T},\dots,\frac{\partial \mathbf{a}_x (\mathbf{\Phi}_x) }{\partial \mathbf{x}_n^T} \right]
    \label{eq:fisher_relative}
\end{equation} 

The $i^{\text{th}}$ element of the Jacobian $[{\partial \mathbf{a}_x (\mathbf{\Phi}) }/{\partial \mathbf{x}_i^T}]$ is given by: 

\begin{equation}
    \left[ \frac{\partial {a}(\mathbf{x}_1,\mathbf{x}_2)^T }{\partial \mathbf{x}_i^T}, \frac{\partial {a}(\mathbf{x}_1,\mathbf{x}_3)^T }{\partial \mathbf{x}_i^T},\dots,\frac{\partial {a}(\mathbf{x}_{n-1},\mathbf{x}_n)^T }{\partial \mathbf{x}_i^T},\right]
\end{equation}

Where $\mathbf{x}_i$ is the position of satellite $i$. $\forall 1 \leq j,k \leq n, j \neq k$ we have:

\begin{equation}
    \frac{\partial {a}(\mathbf{x}_j,\mathbf{x}_k)}{\partial \mathbf{x}_i^T} =
    \begin{cases}
     d_{ik}^{-1}(\mathbf{x}_j-\mathbf{x}_k)^T & \text{if } i = j \\
     -d_{ik}^{-1}(\mathbf{x}_j-\mathbf{x}_k)^T & \text{if } i = k \\
    \mathbf{0}^T & \text{otherwise}
    \end{cases}
\end{equation}

The $1,2,\dots,k$ entries are the unique pairwise links between satellite $i$ and all connected satellites, and as before $d_{ik}$ is the range between nodes $i$ and $k$. The CRB for the anchorless localisation of the Starlink network of satellites is given by substituting the relative Fisher information matrix for an individual satellite (\ref{eq:fisher_relative}) in to (\ref{eq:CRB_RMS}). The code we used to calculate the CRB is available on GitHub \cite{github}.


\subsection{Assumptions on $\gamma$}
The equations presented in the previous sections show that the value of the CRB is highly sensitive to the value of $\gamma$, with larger values of $\gamma$ resulting in smaller CRB values. This implies that the accuracy achievable with cooperative localisation in Starlink will be dependent on the characteristics of the inter-satellite links. Unfortunately, the details of these inter-satellite links are not publicly available, but it is possible to make some initial statements of the link characteristics required for cooperative localisation in Starlink. The expression for $\gamma$ in case of TOA measurements is given by (\ref{eq:gamma}), where the $\sigma_T$ is , \begin{equation}
    \sigma_T \geq \frac{1}{8 \pi^2 B T_s F_c^2 \text{SNR}}
\end{equation} where $B$ is the bandwidth in hertz, $F_c$ is the centre frequency in hertz, $T_s$ is the duration of the signal, and $\text{SNR}$ is the signal-to-noise ratio, ignoring the effects of multi-path communications\footnote{Which is a reasonable assumption to make for satellites in orbit} \cite{patwari_locating_2005}. The value for $\gamma$ used in the simulation of Starlink was $\gamma = 29,8605$. From (\ref{eq:gamma}) and assuming that the velocity of propagation $v_p$ is 3$\cdot$10$^5$ $\text{km}\cdot\text{s}^{-1}$, this means that link should satisfy \begin{equation}
    \frac{1}{B T_s F_c^2 \text{SNR}}\leq 4.81\cdot10^{-7} \text{s} 
\end{equation} 


\subsection{Assumption on System Dynamics}
The equations in Sections \ref{derivation:anchored} and \ref{derivation:anchorless} calculate the RCRB on location estimation without considering system dynamics. Intuitively, modelling the time-varying system dynamics could lead to more accurate localisation. For example, in \cite{Guo_2014} the authors employ an Extended Kalman Filter to determine the orbital positions of satellites performing autonomous navigation using inter-satellite measurements coupled with measurements from a reference satellite. However, in this paper we considered only the instantaneous position of the Starlink satellites to calculate the performance bounds on inter-satellite navigation. 

\section{Simulations}
\label{section:Results}

In this section, we evaluate the localization performance of the satellites in the Startlink network modelled in Section \ref{Methods}, using the lower bounds in Section \ref{calculation}.

\subsection{Anchored Cram$\acute{\text{e}}$r-Rao Bound}
\label{section:anchored_results}
\begin{figure}[t]
    \centering
    \includegraphics[width=0.9\textwidth]{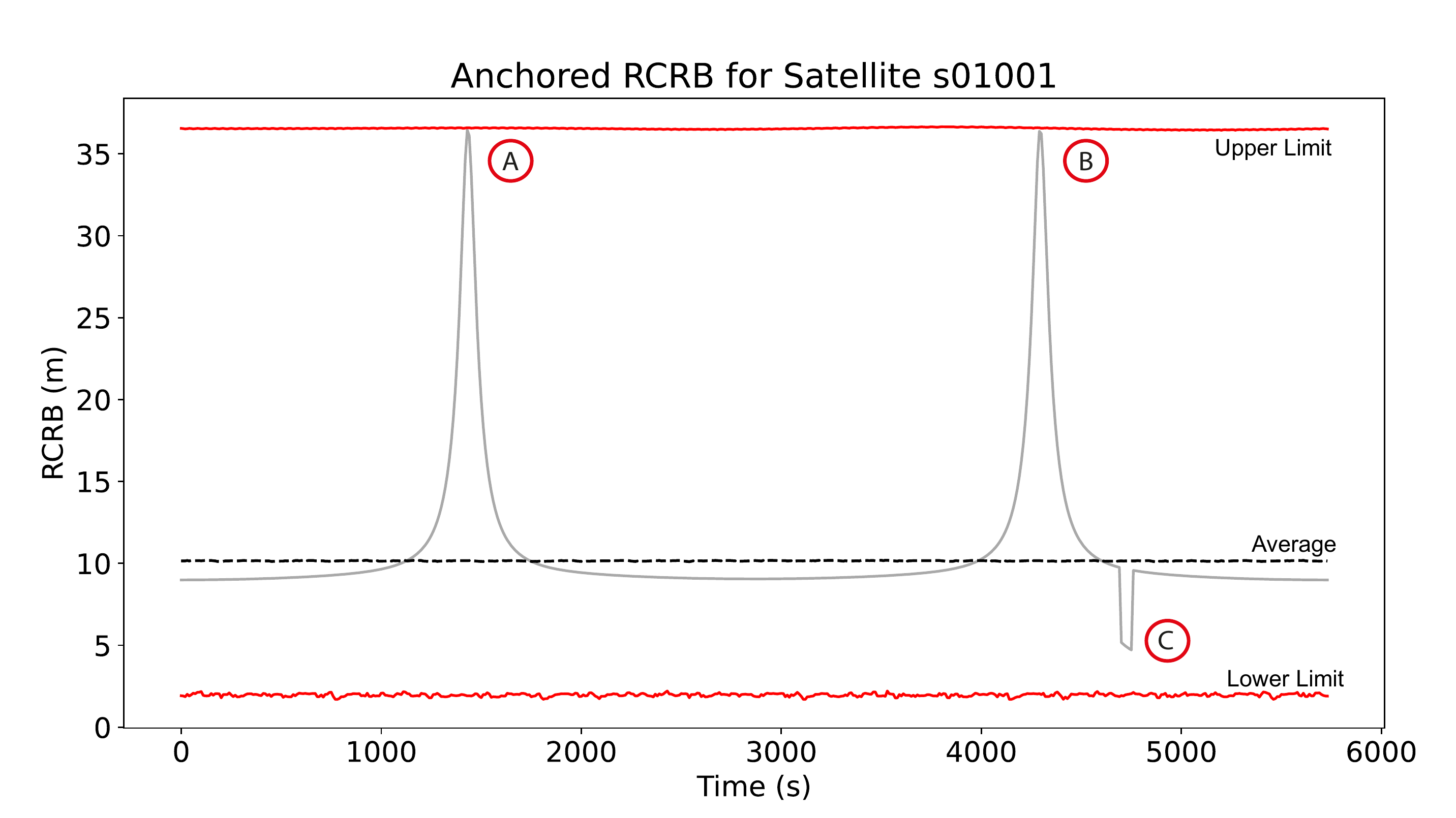}
    \caption[RCRB Results for Starlink]{The average RCRB for Starlink constellation is shown as a dashed black line, and indicates that the Starlink satellites can be localised to within 10.15 metres on average. The upper and lower red lines indicates the maximum (36.5 m) and minimum (~2m) RCRB, respectively. The RCRB for satellite \textit{s01001} is shown in grey, and has prominent peaks (labelled A and B) and a trough labelled C. As shown in Figure \ref{fig:GroundTrack} and \ref{fig:GroundTrack_Small}, these peaks are caused by dilution of precision as the satellite reaches high latitudes and the trough is a result of additional localisation accuracy while the satellite is in range of a ground station. Taking \textit{s01001} as a reference, the results indicate that it is possible to localise satellites to within less than 10 metres for the majority of its orbit. The position of \textit{s01001} is shown in the rendering in Figure \ref{fig:snapshots}.}
    \label{fig:CRLB_results}
\end{figure}

\begin{figure}[t]
  \subfloat[Situation A, $t$=1430 \textit{s}]{
	\begin{minipage}[c][0.9\width]{
	   0.3\textwidth}
	   \centering
	   \includegraphics[width=1\textwidth]{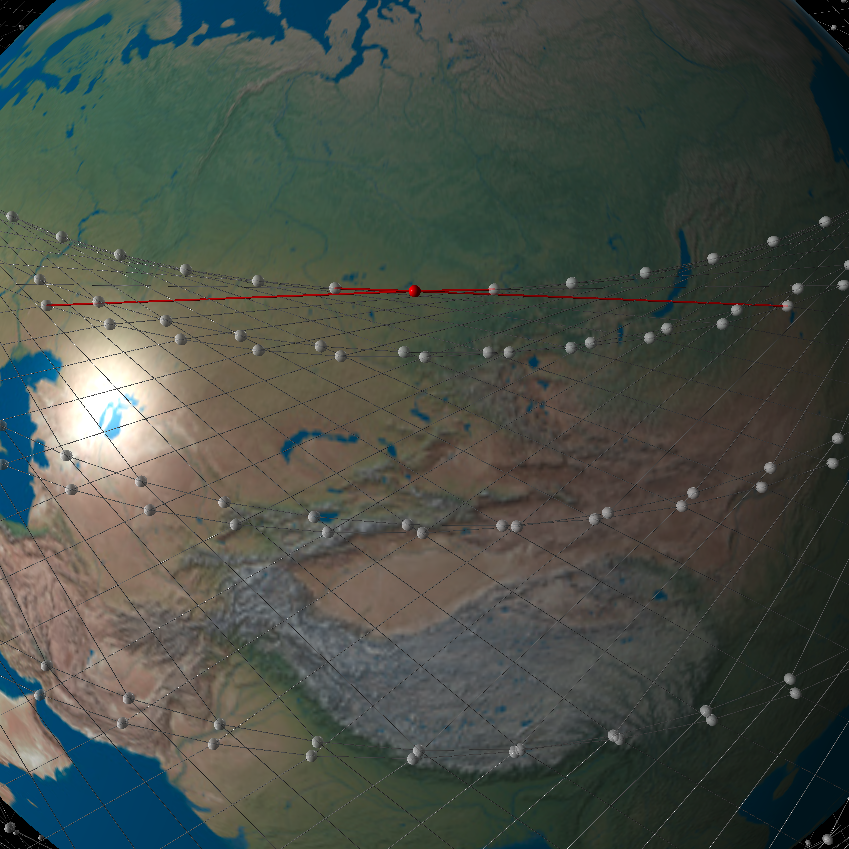}
	\end{minipage}}
 \hfill 	
  \subfloat[Situation B, $t$=4300 \textit{s}]{
	\begin{minipage}[c][0.9\width]{
	   0.3\textwidth}
	   \centering
	   \includegraphics[width=1\textwidth]{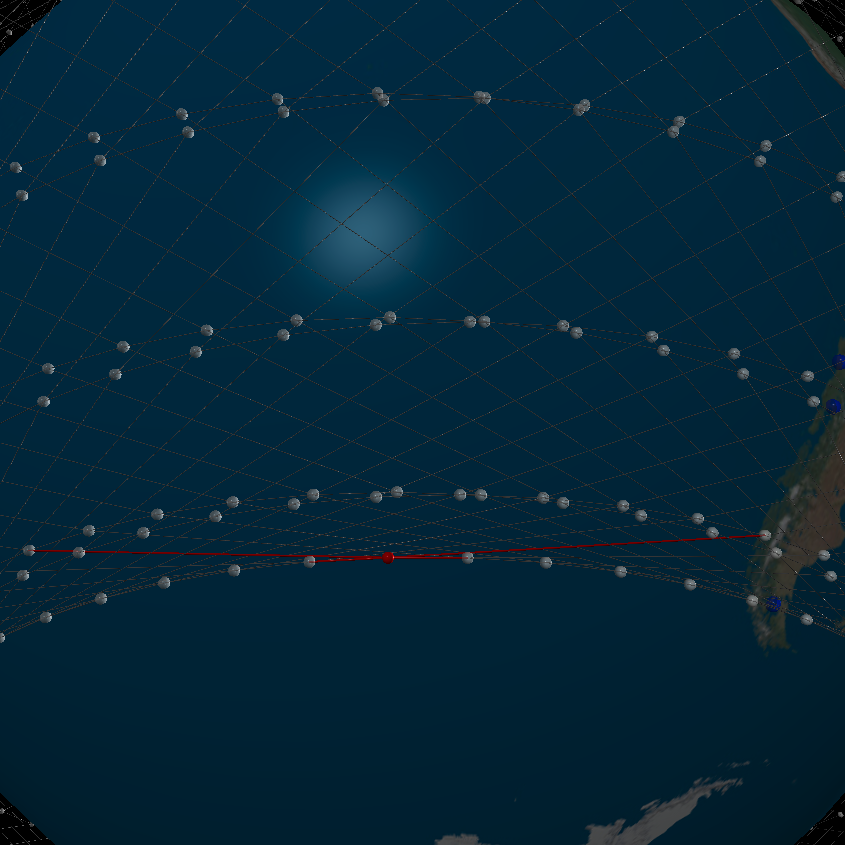}
	\end{minipage}}
 \hfill	
  \subfloat[Situation C, $t$=4750 \textit{s}]{
	\begin{minipage}[c][0.9\width]{
	   0.3\textwidth}
	   \centering
	   \includegraphics[width=1\textwidth]{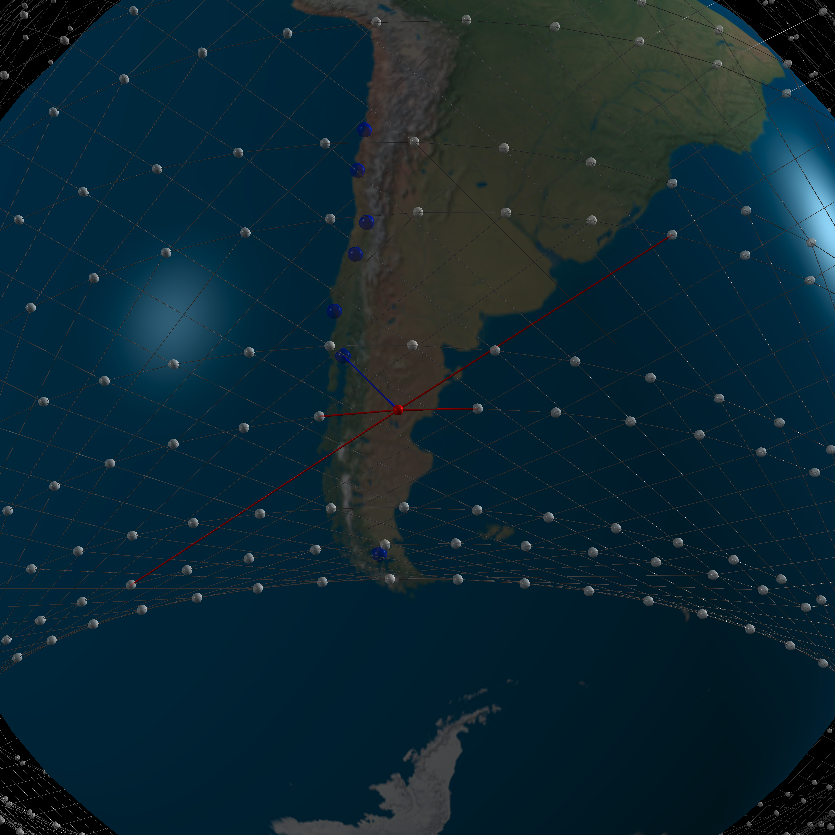}
	\end{minipage}}
\caption[Simulation Snapshots]{Snapshots of the satellite \textit{s01001} in the Starlink simulation at the times highlighted in \ref{fig:CRLB_results}. For situation A, at a time of $t$=1430 \textit{s}, satellite \textit{s01001} is at the highest latitude in its orbit. Situation B, at a time of $t$=4300 \textit{s}, the satellite is at its lowest latitude. Situation C shows satellite \textit{s01001} passing over Tierra del Fuego at the southernmost tip of Chile. In each situation, the connections between satellite \textit{s01001} and other nodes (including ground stations) are shown in red (blue). These three situations correspond to extremes in the CRB for satellite \textit{s01001}.}
\label{fig:snapshots}
\end{figure}

Calculating the instantaneous CRB of all Starlink satellites requires the Fisher matrix to be calculated and inverted for $1584$ satellites at each of $573$ time steps. To obtain a single figure of merit,  (\ref{eq:CRB_RMS}) is used to determine the RCRB. Calculating the RCRB for Starlink during a full orbital period of $T = 5730$ seconds gives the results shown in Figure \ref{fig:CRLB_results}. The average RMSE (Root Mean Square Error) is shown as a dashed black line, and has a fairly constant value of $10.15$m. The value of the CRB varies between a maximum of $36.5$m and a minimum close to $2$m. Figure \ref{fig:CRLB_results} also shows the RCRB for a single satellite (\textit{s01001}) over the course of its orbit. The value is mostly close to the average, but has two prominent peaks at $t=1430$ and $t=4300$ seconds. There is also a noticeable dip in the value at $t=4750$. These three situations (labelled A,B, and C) are rendered in Figure \ref{fig:snapshots}.

\FloatBarrier

\begin{figure}[t]
    \centering
    \includegraphics[width=0.9\textwidth]{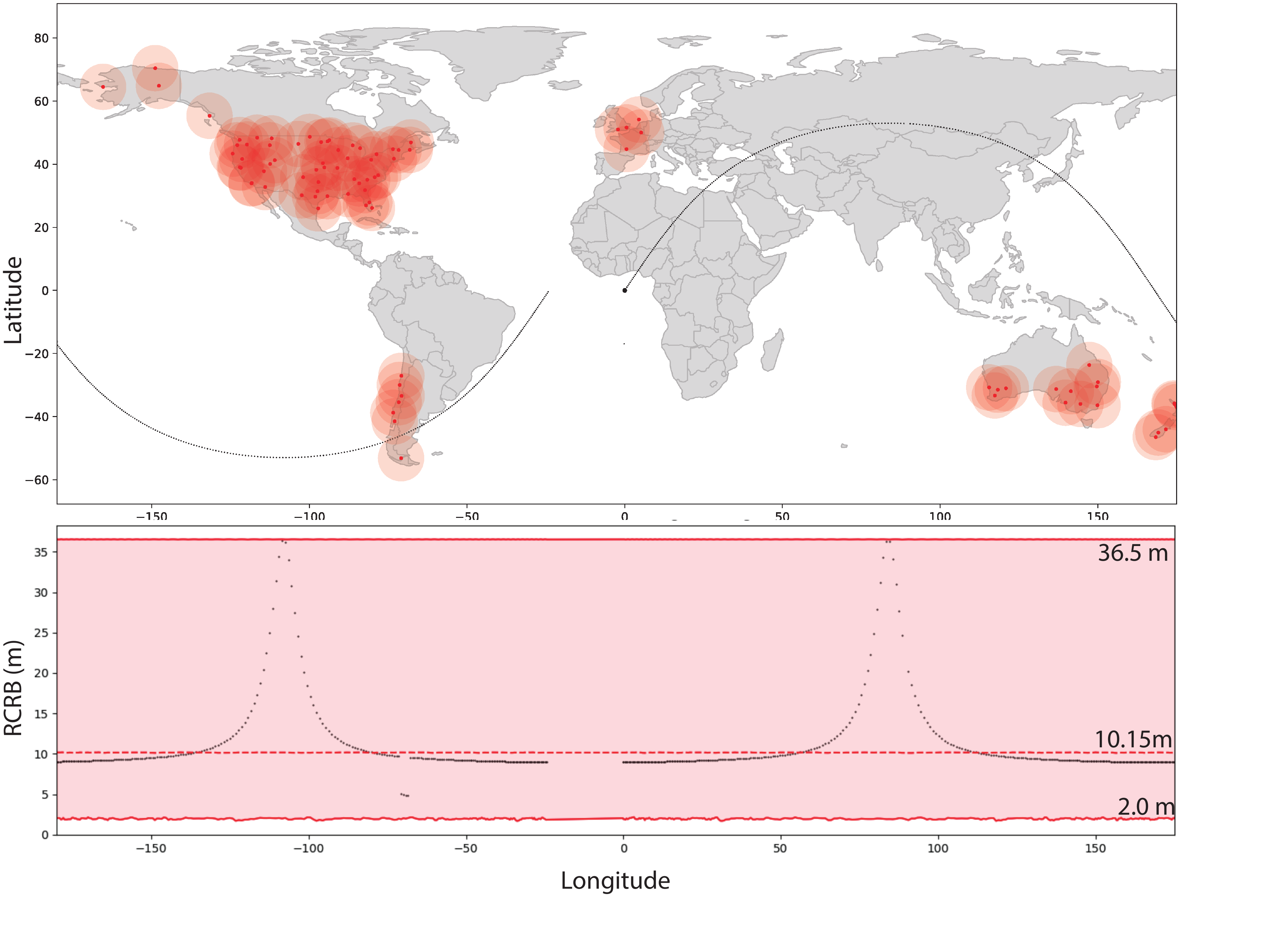}
    \caption[Ground Track for satellite \textit{s01001}]{The upper figure shows the ground track for satellite \textit{s01001} as well as the position of the Starlink ground stations. The lower figure shows the CRB against longitude, with the average CRB for he constellation shown as a dashed red line and the area between the maximum and minimum values for the constellation are shaded in red. Referring to the two plots, it is clear that the peaks in the CRB correspond to the highest and lowest latitudes for \textit{s01001}'s orbit, and that the trough in the CRB occurs when \textit{s01001} is in range of a ground station in South America. The pass of \textit{s01001} over the ground station is shown in detail in Figure \ref{fig:GroundTrack_Small}.}
    \label{fig:GroundTrack}
\end{figure}

\begin{figure}[ht]
    \centering
    \includegraphics[width=0.8\textwidth]{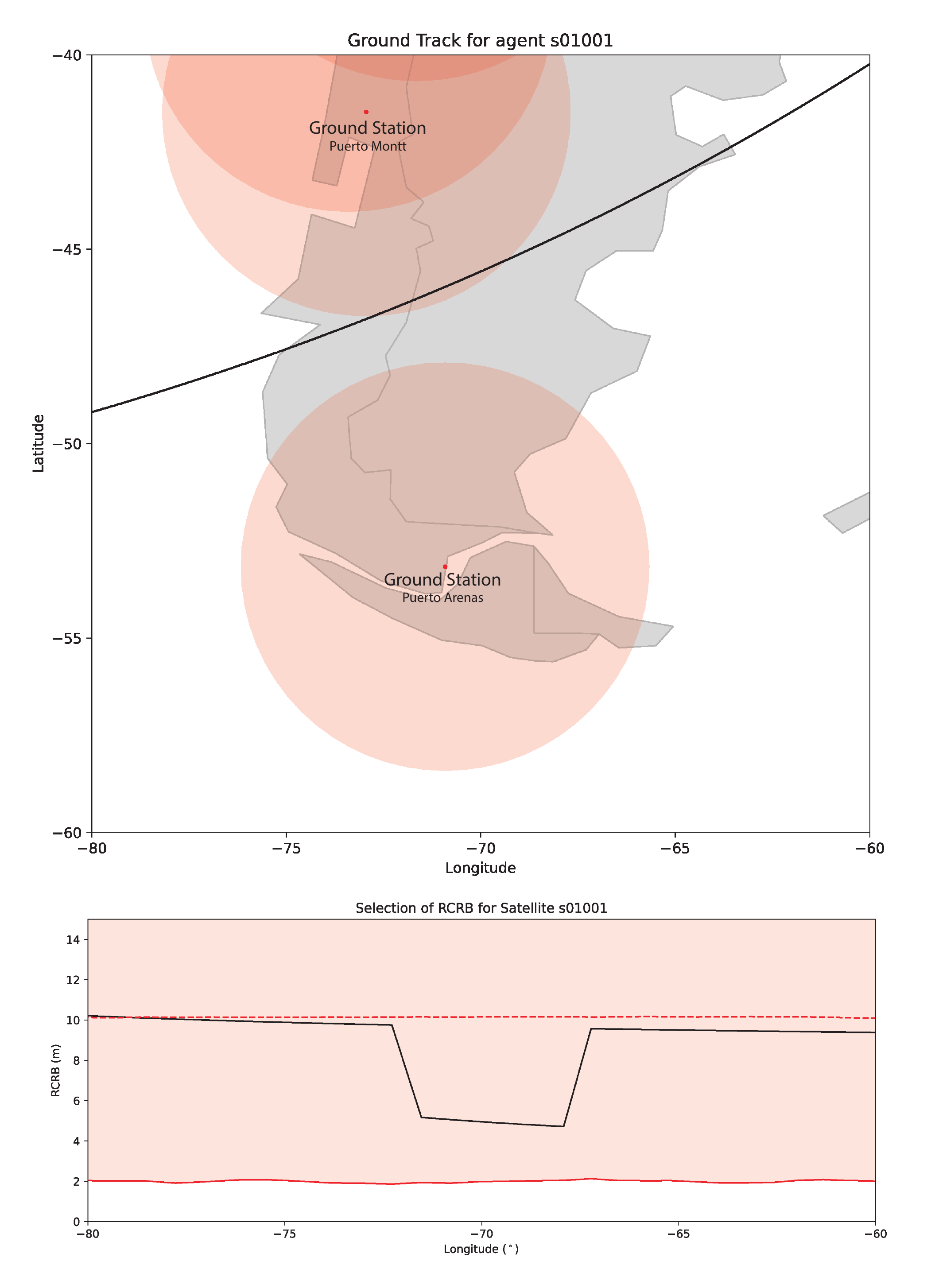}
    \caption[Detailed ground track for \textit{s01001}]{The figure shows satellite \textit{s01001} passing over Tierra del Fuego at the southernmost tip of Chile, as well as the CRB during this pass. Comparison of the two plots shows that the CRB drops by roughly 50\% while it is in range of the ground station at Puerto Montt. The overall trend in the CRB, which is a gradual decrease driven by the geometry of the Starlink network, continues even while \textit{s01011} is in range of the ground station.}
    \label{fig:GroundTrack_Small}
\end{figure}

Inspecting Figure \ref{fig:CRLB_results} and Figure \ref{fig:snapshots}, as well as the ground tracks shown in Figures \ref{fig:GroundTrack} and \ref{fig:GroundTrack_Small} allows us to interpret the peaks and troughs in the CRB for satellite \textit{s01001}. Situations A and B occur when the satellite is at the highest and lowest latitude in its orbit. The two renderings in Figure \ref{fig:snapshots} show why this occurs — the geometrical arrangement of connections with other satellites is less evenly distributed than for the rest of the orbit. This results in an effect similar to dilution of precision in Global Positioning Satellites, where closely aligned satellites results in a lower position accuracy.  Figure \ref{fig:GroundTrack} reveals the reason for the lower RCRB in situation C — as satellite \textit{s01001} passes over a ground station in southern Chile, the connection to the ground station provides more information, reducing the value of the RCRB.

The pass of \textit{s01001} above a ground station is shown in greater detail in Figure \ref{fig:GroundTrack_Small}, which shows the ground track over Tierra del Fuego and a detailed plot of the CRB for \textit{s01001}. The CRB drops by around 50\% as soon as it is within communication range of the ground station at Puerto Montt. While the CRB is reduced by the connection to a ground station, the underlying trend in the CRB is unchanged. This trend is driven by the changing geometry of the Starlink network, and can be seen as the gradual decrease in the plot of \textit{s01001}'s CRB even while the satellite is in range of the Puerto Montt ground station.

The results shown in Figure \ref{fig:CRLB_results} are common to other satellites in Starlink, as shown in Figure \ref{fig:boxplots}, which shows box plots for six Starlink satellites in different orbital planes. Figure \ref{fig:boxplots} confirms that the RCRB remains close to $10.15$m throughout each satellite's orbit, but with occasional peaks as the satellite approaches high latitudes.

\FloatBarrier

\subsection{Anchorless Cram$\acute{\text{e}}$r-Rao Bound}
\label{section:anchorless_results}

\begin{figure}[ht]
    \centering
    \includegraphics[width=0.9\textwidth]{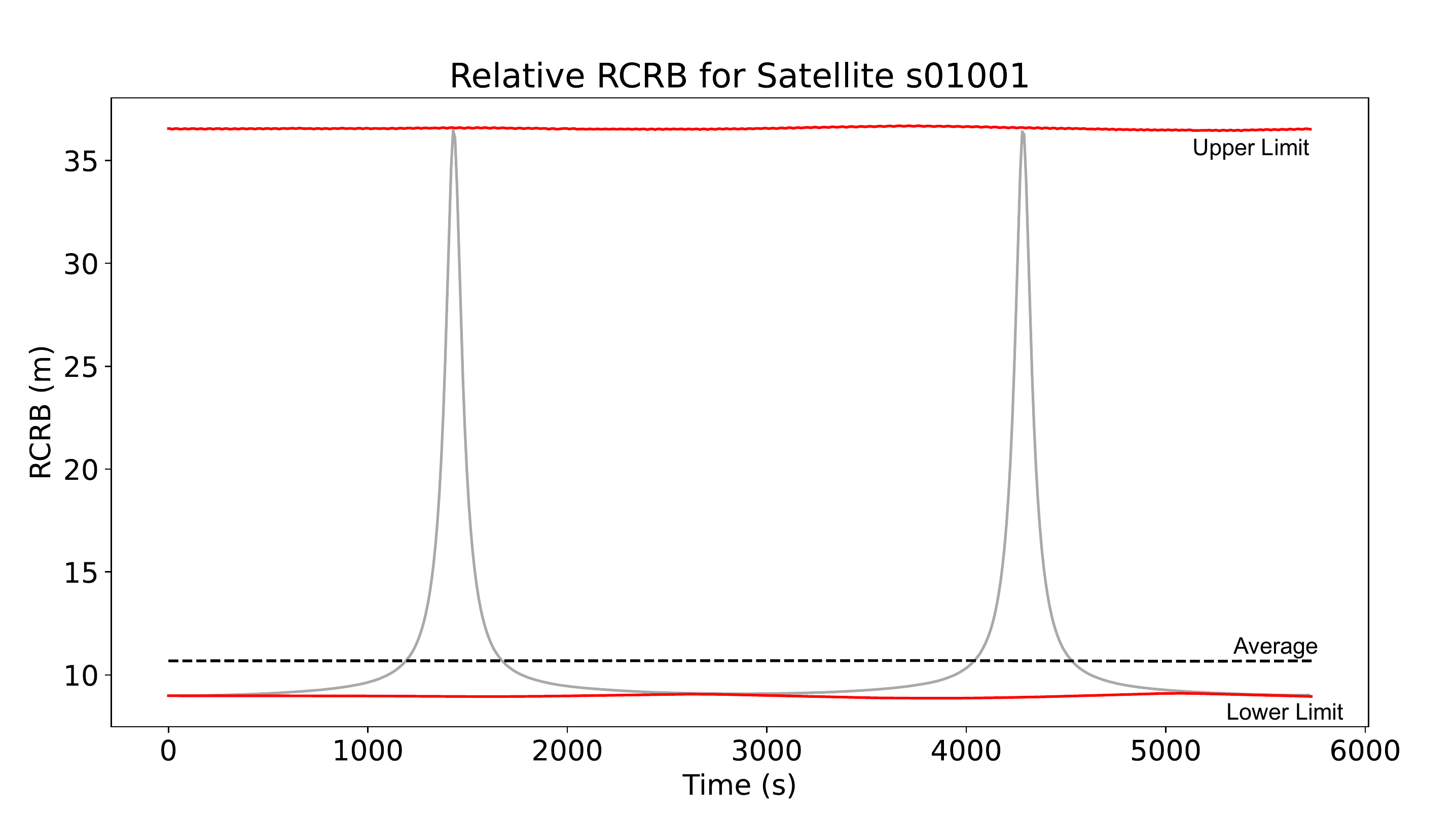}
    \caption[Relative RCRB Results for Starlink]{The average RCRB for the Starlink constellation using only relative positioning is shown as a dashed black line, and indicates that the Starlink satellites can be localised to within 10.68 metres on average, slightly less performant than when ground stations are considered. The shaded red area indicates the area between the maximum (36.64 m) and minimum (8.87 m) values for the RCRB. The RCRB for satellite \textit{s01001} is shown in grey, and resembles the results shown in Figure \ref{fig:CRLB_results}. Taking \textit{s01001} as representative of any satellite in the constellation, the results indicate that it is possible to relatively localise satellites to an accuracy of just over 10 metres for the majority of their orbit.}
    \label{fig:CRLB_relative_results}
\end{figure}

The RCRB for cooperative relative localisation —that is, localisation purely within the constellation and without considering the ground stations— is shown in Figure \ref{fig:CRLB_relative_results}. The RCRB for cooperative localisation for satellite s010001 are similar to those obtained for cooperative localisation with ground stations shown in Figure \ref{fig:CRLB_results}, albeit without the prominent trough during a ground station pass. The RCRB for purely relative localisation shows that this approach is slightly less performant than for localisation including measurements from ground stations, on average 10.68 rather than 10.15 metres. The results still indicate that any satellite in the constellation can be relatively localised  to an accuracy of just over 10 metres for the majority of its orbit. However, it is worth again noting the caveat that these results hold for the relative location of satellites rather than absolute localisation with respect to Earth or other space objects.

\begin{figure}
\begin{subfigure}{\textwidth}
  \centering
  \includegraphics[width=0.9\textwidth]{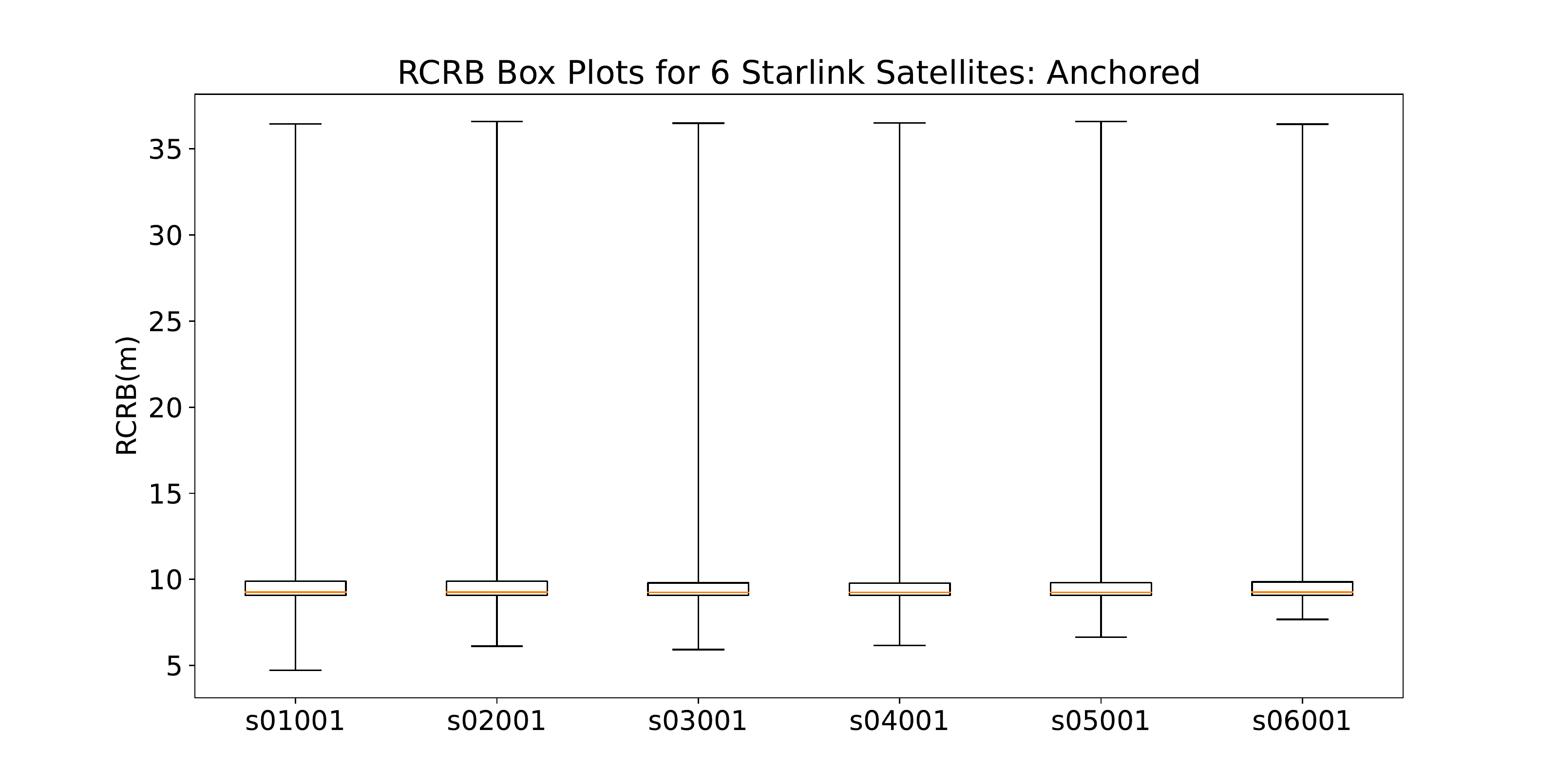}  
  \caption{Box plots showing the RCRB for 6 anchored Starlink satellites}
  \label{fig:sub-first}
\end{subfigure}
\newline

\begin{subfigure}{\textwidth}
  \centering
  \includegraphics[width=0.9\textwidth]{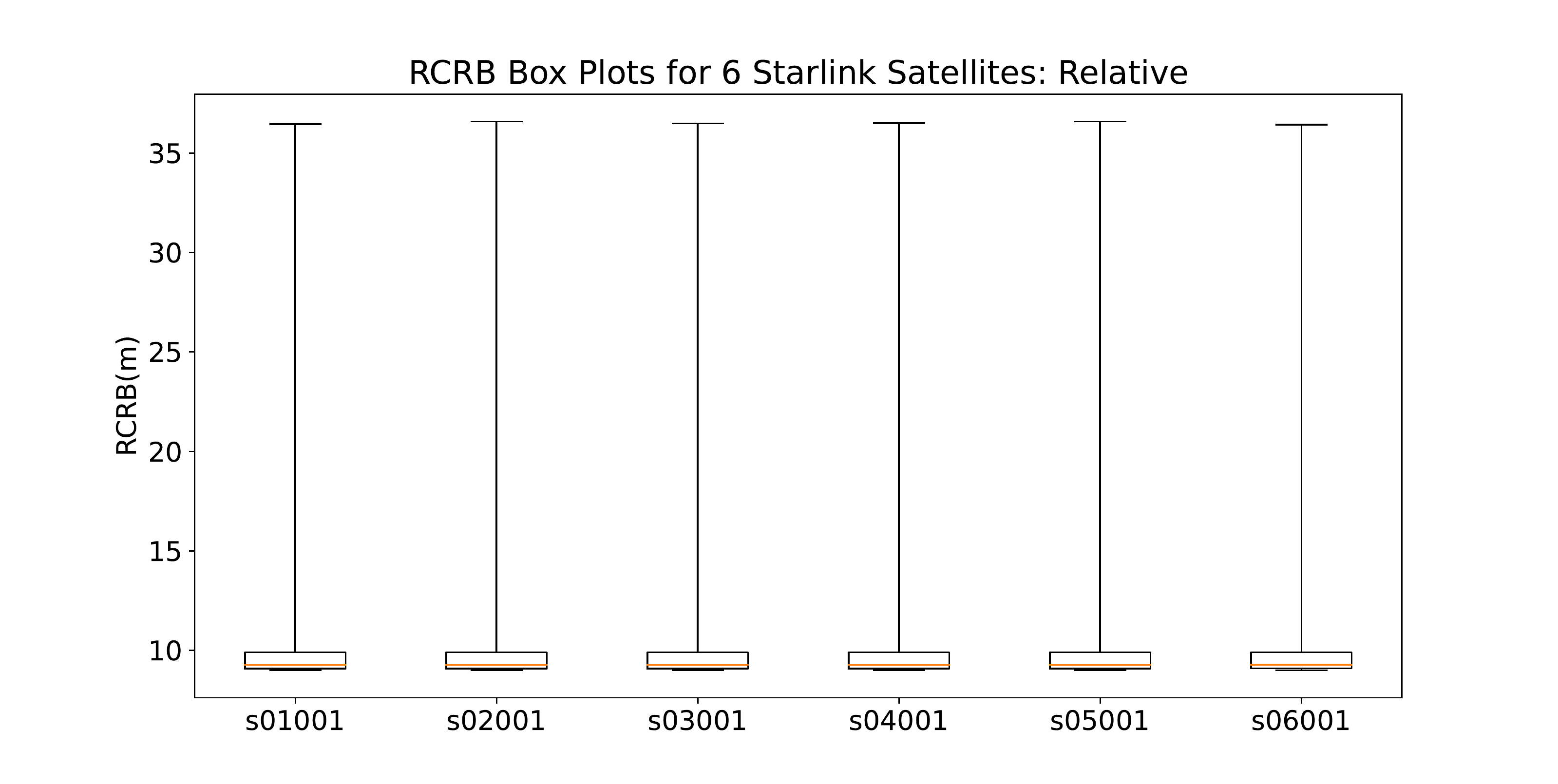}  
  \caption{Box plots showing the RCRB for 6 Starlink satellites using relative navigation}
  \label{fig:sub-third}
\end{subfigure}
\caption{Box plots showing the variability of the RCRB data for six Starlink satellites using both anchored and relative navigation. The six satellites in question are s01001, s02001, s03001, s04001, s05001, s06001 and are from six different orbital planes in the megaconstellation. The upper figure shows the results for anchored navigation (\textit{i.e.} including ground stations) and the lower figure shows the results for relative navigation. The yellow lines show the median results, the box shows the inter-quartile range, and the whiskers extend to the minimum and maximum values. The figure demonstrates that the performance of both relative and anchored navigation mainly clusters around 10 metres, with large outliers extending to roughly 35 metres. As described in Figure \ref{fig:GroundTrack}, this is due to dilution of precision while the satellite is at high latitudes. The figure also demonstrates that there is little satellite-to-satellite variation in cooperative localisation performance.}
\label{fig:boxplots}
\end{figure}

The plots in Figure \ref{fig:boxplots} confirm that relative localisation is similarly performant to localisation also using ground stations for the majority of a satellite's orbit. The lower average RCRB when also considering ground stations is to due to improve localisation performance when Starlink satellites are in range of a ground station. In effect, multiple troughs such as the one shown in Figure \ref{fig:GroundTrack_Small} reduce the overall average RCRB for cooperative localisation augmented by ground stations. Comparing the two sets of box plots in Figure \ref{fig:boxplots}, however, shows that the performance in both cases is almost identical for the majority of the orbit.

\subsection{Discussion}
\label{section:Discussions}
The results indicate that the position of Starlink satellites can be determined from inter-satellite measurements to an average RMSE of approximately 10.15 metres for the majority of their orbit using ground stations to aid localisation, and 10.68 metres on average when performing only relative localisation. These results are similar to those reported for other constellations using cooperative inter-satellite navigation \cite{dave2020autonomous} but also show room for improvement. However these results are highly dependant on the value of $\gamma$ used to calculate the CRB and also ignore the dynamics of the system. Our results could potentially be improved by considering the orbital dynamics of the satellites in Starlink, for example by combining intersatellite cooperative navigation with an Extended Kalman Filter. In future work we aim to explore how this affects our results. Furthermore, as discussed in Section \ref{section:crb}, the RCRB is highly dependent on $\gamma$, which is determined bu the characteristics of the intersatellite links. Repeating the analysis for a range of link characteristics based on existing satellite hardware could allow a technical trade-off to be performed. Other aspects of the inter-satellite links, such as equipment duty cycles, could also affect inter-satellite links — for example, in \cite{dave2020autonomous}, the authors considered the duty cycle of communications in satellite pairs and small satellite constellations.

\section{Conclusions}
\label{section:Conclusions}
In this paper, we presented a Phase-1 model of the Starlink network and investigated the potential of cooperative localisation of the Starlink satellites by studying the performance of unbiased estimators for anchored and anchorless scenarios. The results of our research on cooperative localisation in Starlink show that localisation determined from inter-satellite measurements and ground stations can have at best an an average RMSE of approximately $10.15$ metres over the majority of a satellite's orbit, which could improve space situational awareness and provide a redundant way to localise swarm satellites in orbit. Relative localisation using only inter-satellite measurements has a slightly poorer performance with an average RMSE of $10.68$ metres. The results also show that inter-satellite cooperative localisation is dependent on the characteristics of the constellation's time-varying geometry and the characteristics of inter-satellite links. In this work, we considered the performance of instantaneous positioning of the Starlink satellites. However, to emulate a more realistic scenarios, Bayesian CRBs can be computed for time-varying position estimation, which is beyond the scope of this work, and will be addressed in future research.

\bibliographystyle{elsarticle-num} 
\bibliography{Bibliography}

\end{document}